\documentclass[onecolumn,unpublished,a4paper]{quantumarticle}
\usepackage[utf8]{inputenc}
\usepackage[backend=biber,style=alphabetic,maxbibnames=999,backref=true]{biblatex}
\usepackage{amsmath}
\usepackage{amssymb}
\usepackage{amsthm}
\usepackage{amsfonts}
\usepackage{changepage}
\usepackage[caption=false]{subfig}
\usepackage[colorlinks]{hyperref}
\usepackage[all]{hypcap}
\usepackage{import}
\usepackage{tikz}
\usepackage{relsize}
\usepackage{color,soul}
\usepackage{capt-of}
\usepackage{mathtools}
\usepackage{float}
\usepackage[section]{placeins}
\usepackage{mdframed}
\usepackage{listings}
\usepackage[T1]{fontenc}  
\usetikzlibrary{decorations.pathreplacing}
\usepackage{braket}

\DefineBibliographyStrings{english}{
  backrefpage  = {Cited on page},
  backrefpages = {Cited on pages}
}
\renewcommand\thesection{\arabic{section}}

\theoremstyle{definition}

\theoremstyle{definition}

\theoremstyle{definition}

\newcommand{\eq}[1]{\hyperref[eq:#1]{Equation~\ref*{eq:#1}}}
\renewcommand{\sec}[1]{\hyperref[sec:#1]{Section~\ref*{sec:#1}}}
\DeclareRobustCommand{\app}[1]{\hyperref[app:#1]{Appendix~\ref*{app:#1}}}
\newcommand{\fig}[1]{\hyperref[fig:#1]{Figure~\ref*{fig:#1}}}
\newcommand{\tbl}[1]{\hyperref[tbl:#1]{Table~\ref*{tbl:#1}}}
\newcommand{\theoremref}[1]{\hyperref[theorem:#1]{Theorem~\ref*{theorem:#1}}}
\newcommand{\definitionref}[1]{\hyperref[definition:#1]{Definition~\ref*{definition:#1}}}

\DeclareFixedFont{\ttb}{T1}{txtt}{bx}{n}{8}
\DeclareFixedFont{\ttm}{T1}{txtt}{m}{n}{8}
\definecolor{deepblue}{rgb}{0,0,0.5}
\definecolor{deepred}{rgb}{0.6,0,0}
\definecolor{deepgreen}{rgb}{0,0.5,0}

\begin{document}
\title{
A Classical-Quantum Adder with Constant Workspace and Linear Gates
}

\date{\today}
\author{Craig Gidney}
\email{craig.gidney@gmail.com}
\affiliation{Google Quantum AI, Santa Barbara, California 93117, USA}

\begin{abstract}
In 2004, Cuccaro et al found a quantum-quantum adder with $O(n)$ gate cost and $O(1)$ ancilla qubits.
Since then, it's been an open question whether classical-quantum adders can achieve the same asymptotic complexity.
These costs are particularly relevant to modular arithmetic circuits, which often offset by the classically known modulus.

In this paper, I construct an adder that uses 3 clean ancillae and $4n \pm O(1)$ Toffoli gates to add a classical offset into a quantum register.
I also present an adder with a Toffoli cost of $3n \pm O(1)$ that uses 2 clean ancillae and $n-2$ dirty ancillae.
I further show that applying the presented adders conditioned on a control qubit requires no additional workspace or Toffolis.
\end{abstract}

\textbf{Data availability}: \emph{Code and assets created for this paper are available \href{https://zenodo.org/records/15866587}{on Zenodo}~\cite{gidneyzenodo2025constadd}}.

{
  \renewcommand{\contentsname}{}
  \vspace{-1cm}

  \hypersetup{linkcolor=blue}
  \tableofcontents
}

\section{Introduction}
\label{sec:introduction}

The first quantum adder circuits needed $\Theta(n)$ gates and $\Theta(n)$ workspace qubits~\cite{vedral1996arithmetic}.
In 2004, Cuccaro et al found a way to eliminate the workspace, while preserving the $\Theta(n)$ gate count, by temporarily moving carry information into the qubits of the offset register~\cite{cuccaro2004adder}.
However, this can't work for classical-quantum adders where the offset is stored in bits instead of qubits.
Extending the Cuccaro result to the classical-quantum case has remained an open problem since 2004.

There was some progress on the problem, between 2004 and now.
In 2015, I found a solution for the special case where the offset is 1~\cite{gidney2015largeincrement}.
In 2016, Häner et al found a divide-and-conquer classical-quantum adder that used 1 dirty qubit and $\Theta(n \lg n)$ gates~\cite{haner2016factoring}.
But reaching the asymptotic limit of $\Theta(n)$ gates remained unsolved.

In this paper I finally close the remaining asymptotic gap, achieving an $\Theta(n)$ gate count under the constraint of $O(1)$ ancilla qubits.
The new construction is based on three key ideas:

\begin{enumerate}
\item
The ripple carry process of an addition can be streamed with constant workspace by eagerly ``venting'' carry qubits (measuring them in the X basis as part of a measurement based uncomputation~\cite{gidney2019spookypebble}).
This venting process leaves behind phasing tasks that correspond to applying a Z gate to some of the carries of the addition.
\item
The vented phasing tasks can be resolved later by interleaving phase flips and Häner et al's construction for xoring carries into a secondary register~\cite{haner2016factoring}.
\item
Dirty workspace (e.g. the secondary register targeted by the carry-xor) can be provided by splitting the addition into two halves, with each half borrowing the other as dirty workspace.
\end{enumerate}

The paper is structured as follows.
In \sec{construction}, I explain the construction of the adder and some variants.
In \sec{conclusion}, I summarize the results.

\section{Construction}
\label{sec:construction}

\subsection{Notation}

The bit at index $k$ of an integer $x$ is referred to as $x_k$ (for example $x_0$ is the even-vs-odd bit):

\begin{equation}
    x_k = \lfloor x / 2^k \rfloor \bmod 2
\end{equation}

The bitwise xor of $x$ and $y$ is written $x \oplus y$:

\begin{equation}
    x \oplus y = \sum_{k=0}^\infty (x_k + y_k - 2 x_k y_k) 2^k
\end{equation}

(Use the 2-adic metric to get the above sum to converge when $x$ or $y$ are negative.)

The bitwise complement of $x$ is written $\sim x$:

\begin{equation}
    \sim x = -1 \oplus x = -1-x
\end{equation}

The majority of three bits is the most common bit value among the three:

\begin{equation}
\begin{aligned}    
    \text{maj}(a, b, c)
    &= \lfloor (a + b + c) / 2 \rfloor
    \\&= ((a \oplus c) \cdot (b \oplus c)) \oplus c
\end{aligned}
\end{equation}

The ``carry'' of an addition is the extra bit flips that occur due to propagating carries (including from the carry input bit $c_0$).
It's the xor between the actual result of the addition and the xor of its two inputs:

\begin{equation}
    \text{carry}(x, d, c_0) = x \oplus d \oplus (x + d + c_0)
\end{equation}

\begin{figure}
    \centering
    \includegraphics[width=1.0\linewidth]{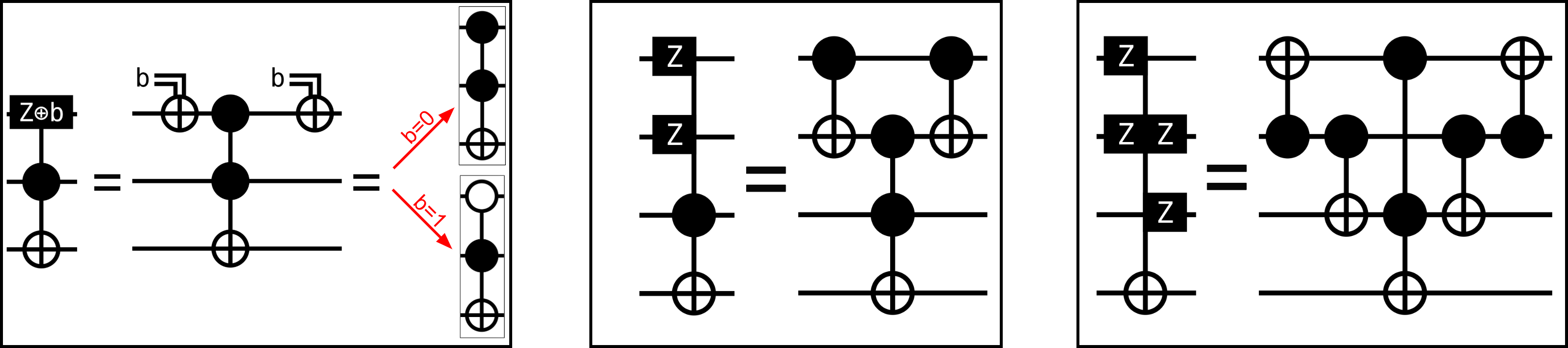}
    \caption{
        Non-standard circuit notation in this paper.
        Left: Bit-inverted controls. A black box containing the text "$Z \oplus b$" means the control's required value is inverted when the bit $b$ is True.
        Middle: Controlling with a ZZ operator.
        Conditioning on a multi-qubit parity operator, like $ZZ$, is shown as black boxes to the side of the control line indicating each of the $Z$ terms.
        Right: Controlling with the intersection of two ZZ operators.
        When an operation has two parity controls, terms from one are placed to the left of the control line while terms from the other are placed to the right.
        I prefer parity controls over CNOT gates because accessing $ZZ$ is notably cheaper than doing CNOTs when compiling into lattice surgery.
    }
    \label{fig:notation}
\end{figure}

Lastly, this paper uses some non-standard circuit notation defined in \fig{notation}.

\subsection{Streaming Addition by Venting Carries}

When performing an addition

\begin{equation}
x^\prime = x + d + c_0
\end{equation}

the first carry bit $c_0$ is given as an input, and each subsequent carry bit $c_k$ is the majority of the previous carry bit and bits from the two values being added:

\begin{equation}
    c_{k+1} = \text{maj}(x_{k}, d_{k}, c_{k})
\end{equation}

After $c_k$ is computed, a bit of the sum $x^\prime_k = x_k \oplus d_k \oplus c_k$ can be computed.
Then one of the input bits can be uncomputed (because $x_k = x^\prime_k \oplus d_k \oplus c_k$) and the next carry bit $c_{k+1}$ can be computed.
At this point $c_k$ becomes unwanted garbage.
Getting rid of this carry garbage is a key obstacle in most quantum adder designs.

An important property that carries satisfy is that the carries produced when computing a sum are reproduced by adding the same offset into the bitwise complement of the sum:

\begin{equation}
\label{eq:carryneg}
\begin{aligned}
    &\text{carry}(\sim x^\prime, d, c_0)
    \\&= \Big( \sim x^\prime \Big) \oplus d \oplus \Big(\sim x^\prime + d + c_0\Big)
    \\&= \Big( \sim(x + d + c_0)\Big) \oplus d \oplus \Big(\sim(x + d + c_0) + d + c_0\Big)
    \\&= \Big(\sim (\text{carry}(x, d, c_0) \oplus x \oplus d)\Big) \oplus d \oplus \Big(-1-(x + d + c_0) + d + c_0\Big)
    \\&= \Big(-1 \oplus \text{carry}(x, d, c_0) \oplus x \oplus d\Big) \oplus d \oplus \Big(-1-x\Big)
    \\&= -1 \oplus \text{carry}(x, d, c_0) \oplus x \oplus d \oplus d \oplus -1 \oplus x
    \\&= \text{carry}(x, d, c_0)
\end{aligned}
\end{equation}

Because of \eq{carryneg}, each carry qubit is a Z-basis function of values available after the addition completes (the carries are ``Z-redundant'').
Measuring a Z-redundant qubit $q$ in the X basis produces a 50/50 random result, where the $|+\rangle$ result means $q$'s value was deleted with no consequences and the $|-\rangle$ result means $q$'s value was deleted but an unwanted phase flip $Z_q$ occurred just before the deletion~\cite{gidney2019spookypebble}.
The deletion is completed by eventually correcting the unwanted phase flip.
Therefore, one way to get carry garbage out of qubits is by converting the garbage into unwanted phase flips by measuring the qubits in the X basis.
Throughout the paper I refer to this conversion process, from Z-redundant qubits into unwanted phasing tasks, as ``venting''.

\begin{figure}
    \centering
    \includegraphics[width=1.0\linewidth]{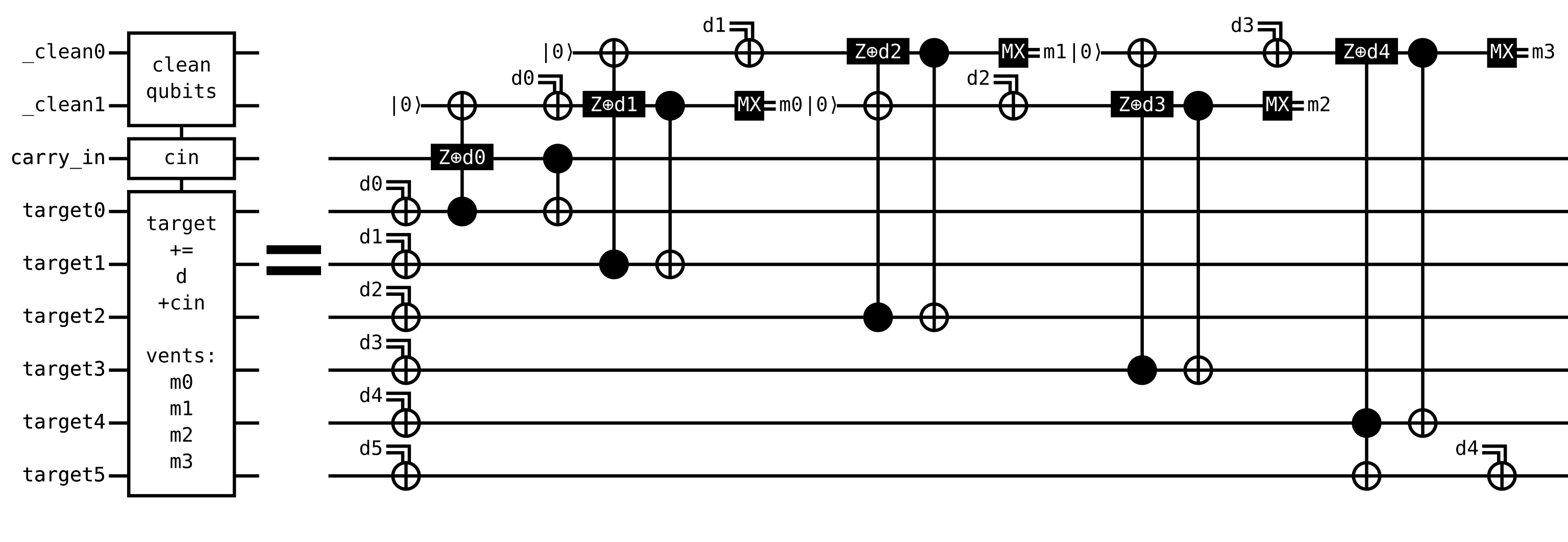}
    \caption{
        An $n=6$ example of an adder that vents carries in order to only use 2 clean ancilla and $n \pm O(1)$ Toffolis.
        Performs $\text{target} \rightarrow (\text{target} + d + \text{carry}_\text{in}) \bmod 2^n$, except that completing the addition requires performing phase flips based on the vented values $m_0, \dots, m_{n-3}$.
        If $m_k$ is true, then a phase flip by $\text{carry}(\sim \text{target}^\prime, d, \text{carry}_\text{in})_{k+1}$ is needed, where $\text{target}^\prime$ is the value of the target register after the addition.
        \protect\input{url-2-clean-vent.tex}
    }
    \label{fig:streaming-adder}
\end{figure}

An adder based on venting carries is shown in \fig{streaming-adder}.
This adder leaves behind unwanted phasing tasks, but only requires two clean ancilla qubits of workspace.

\subsection{Phasing by Carries}

\begin{figure}
    \centering
    \includegraphics[width=1.0\linewidth]{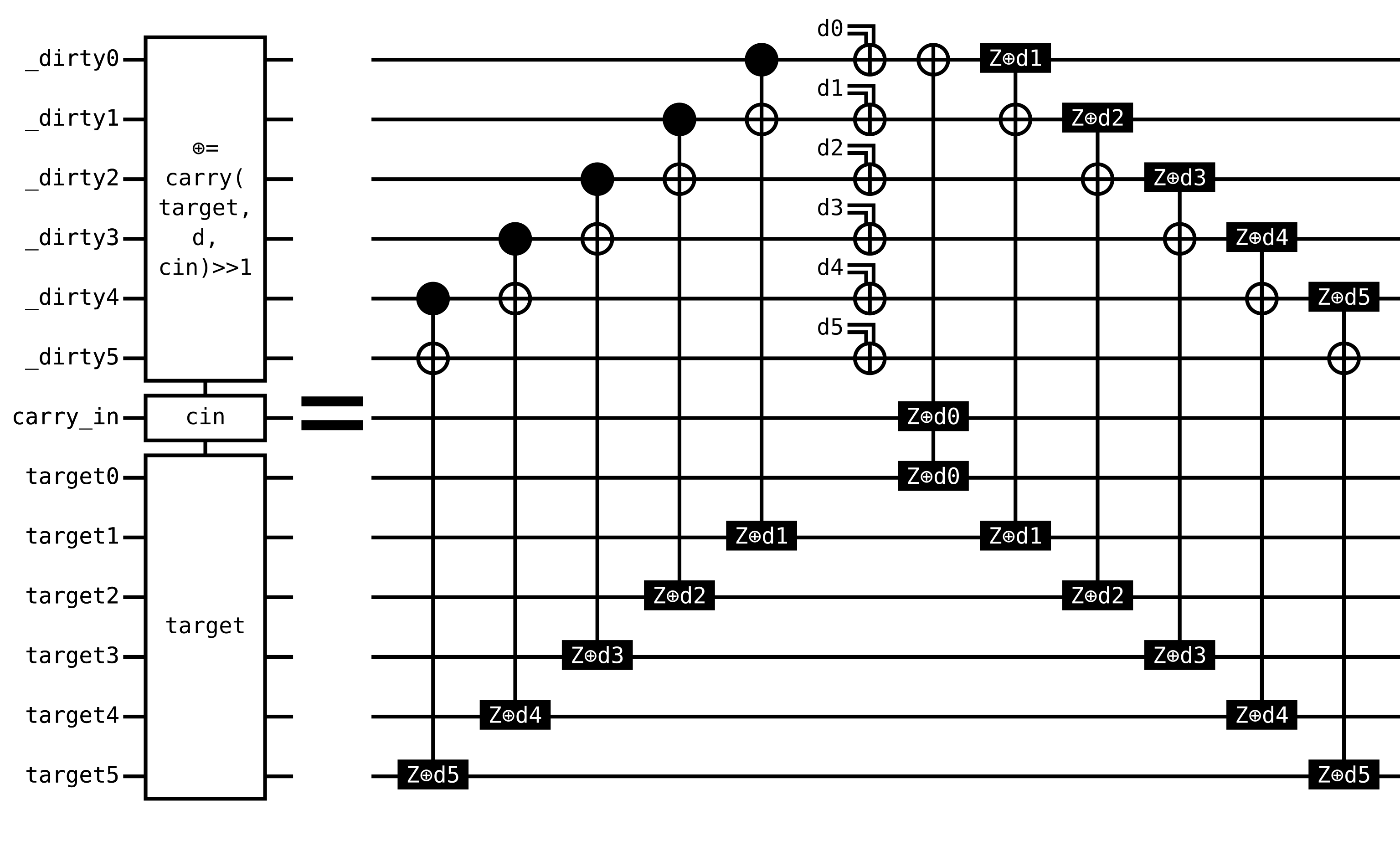}
    \caption{
        An $n=6$ variant of the carry-xor construction from \cite{haner2016factoring}.
        Performs $\text{\_dirty} \rightarrow \text{\_dirty} \oplus \lfloor \text{carry}(\text{target}, d, \text{carry}_\text{in}) / 2 \rfloor$.
        Skips over the least significant carry because that's equal to the carry input and so is trivial to access. \protect\input{url-carry-xor.tex}
    }
    \label{fig:carry-xor}
\end{figure}

In \cite{haner2016factoring}, Häner et al describe a ``carry-xor'' circuit that performs $g \rightarrow g \oplus \text{carry}(x, d, c_0)$.
It xors the carries that would be generated by a planned addition $x \rightarrow x + d + c_0$ into a secondary register $g$.
A minor variation on their construction is shown in \fig{carry-xor}.

It's well known that interleaving phase flips of $q$ with controlled bit flips of $q$ causes a phase flip of whatever is controlling the bit flips.
Applying $Z_q$ then $CX_{c \rightarrow q}$ then $Z_q$ then $CX_{c \rightarrow q}$ is equivalent to applying $Z_c$.
Therefore, because the carry-xor circuit bit flips qubits controlled by the carries of an addition, it's possible to phase flip by the carries of an addition by interleaving phase flips and carry-xors.

\begin{figure}
    \begin{adjustwidth}{-2.5cm}{-2.5cm}
    \centering
    \includegraphics[width=0.7\linewidth]{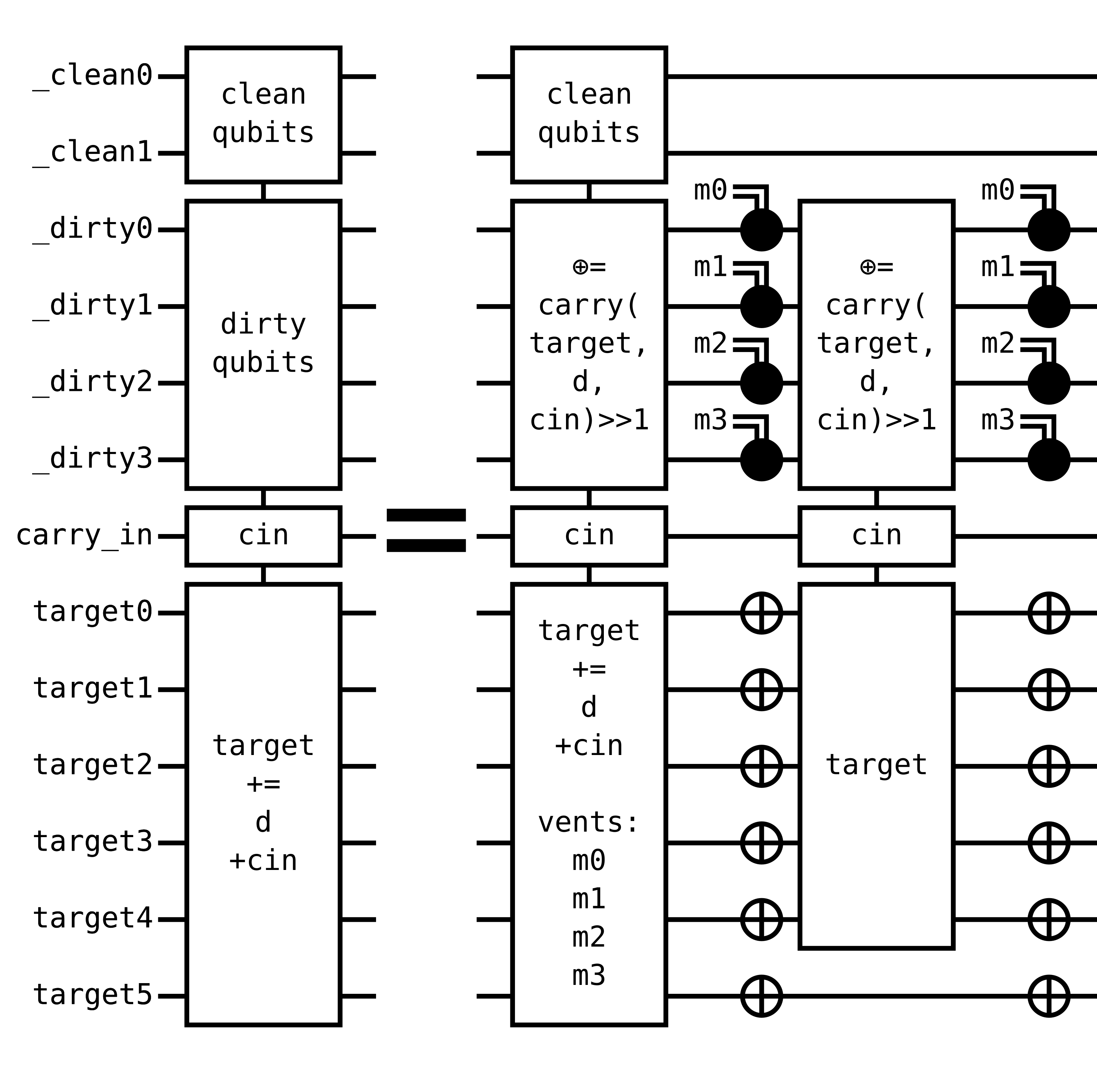}
    \includegraphics[width=1.0\linewidth]{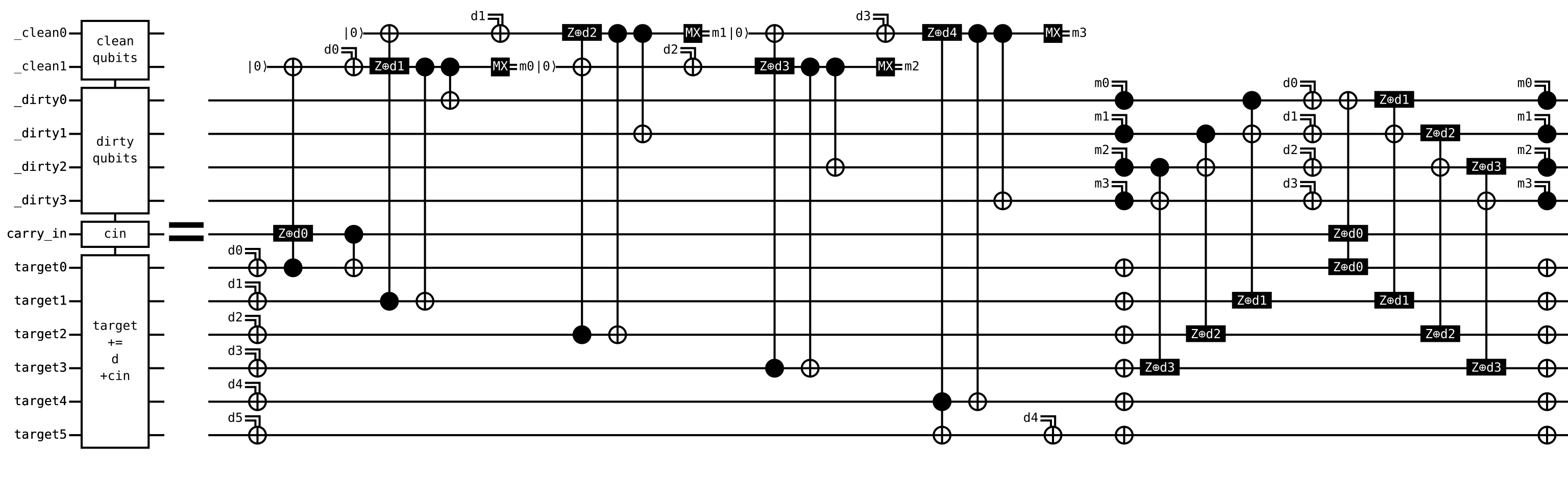}
    \end{adjustwidth}
    \caption{
        An $n=6$ example of an adder that uses 2 clean ancillae, $n-2$ dirty ancillae, and $3n \pm O(1)$ Toffolis.
        Top: decomposed into other subroutines defined by the paper.
        Bottom: further decomposed into constant sized gates.
        Performs $\text{target} \rightarrow (\text{target} + d + \text{carry}_\text{in}) \bmod 2^n$.
        The circuit ends with an instance of the carry-xor from \fig{carry-xor}, conjugated by classically-controlled Z gates to correct the phase vented by the streaming adder from \fig{streaming-adder} at the start of the circuit.
        Note that the streaming adder is tweaked to xor carries into the dirty qubits as it runs, saving one instance of the expensive carry-xor construction.
    }
    \label{fig:add-2-clean-n-dirty}
\end{figure}

To avoid leaving $g$ perturbed, two carry-xors are needed.
The first carry-xor can be achieved cheaply by using the carry qubits that were already being computed during the addition.
The second carry-xor is achieved using the Häner et al construction.
Combined with some classically controlled $Z$ gates, and the vented adder described in the previous subsection, the result is an adder that uses $3n \pm O(1)$ Toffolis, $n-2$ dirty qubits, and 2 clean qubits.
An example is shown in \fig{add-2-clean-n-dirty}.

\begin{figure}
    \begin{adjustwidth}{-2.5cm}{-2.5cm}
    \centering
    \includegraphics[width=0.9\linewidth]{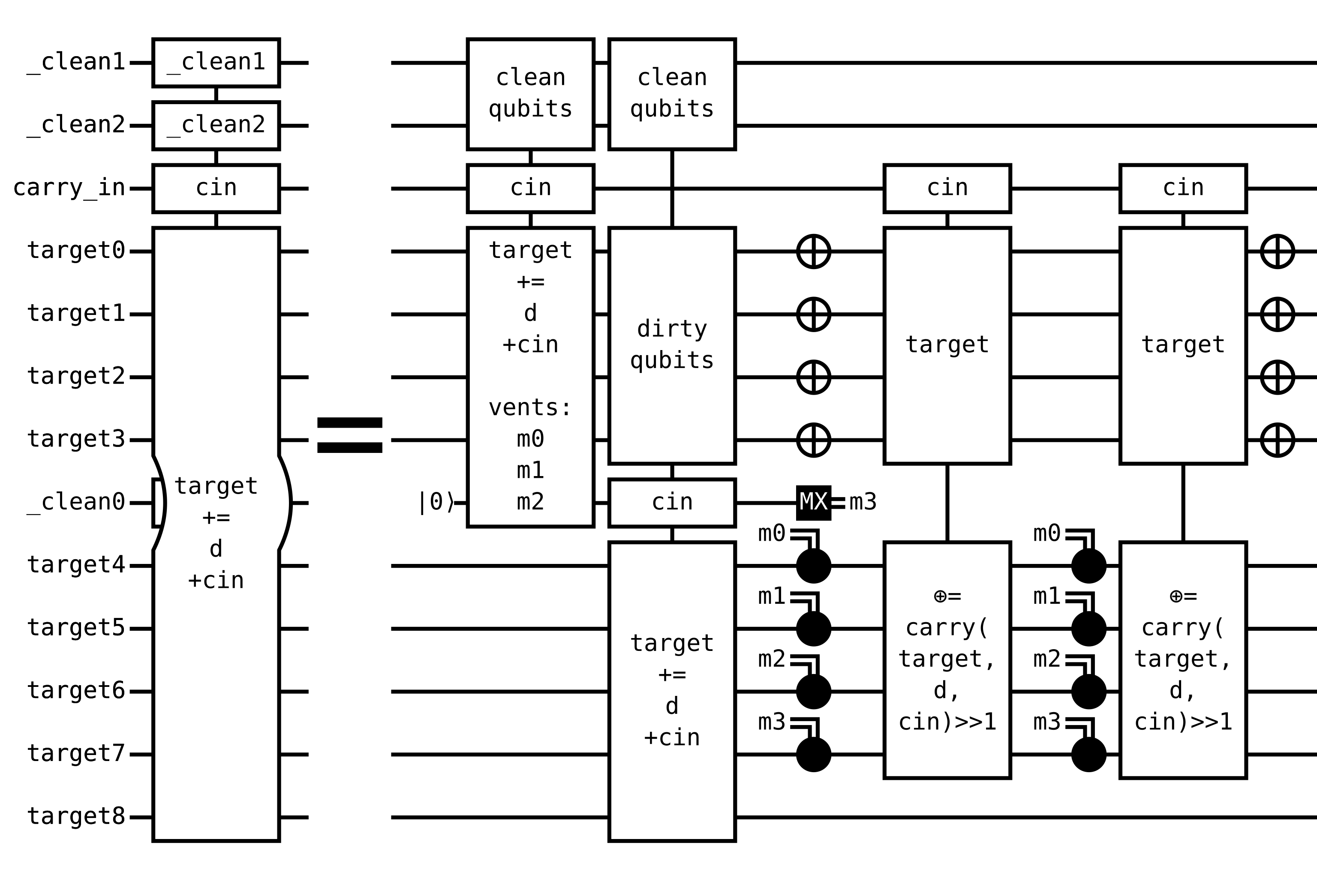}
    \includegraphics[width=1.0\linewidth]{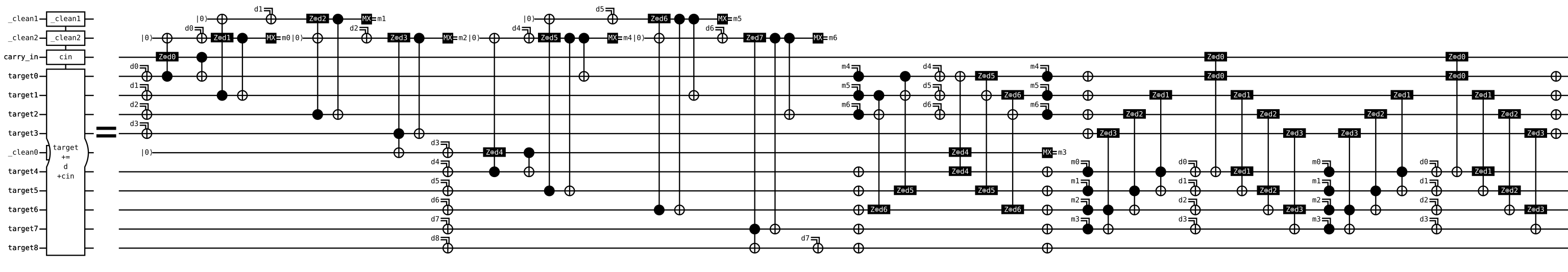}
    \end{adjustwidth}
    \caption{
        An $n=9$ example of an adder circuit that uses 3 clean ancillae and $4n \pm O(1)$ Toffolis.
        Top: decomposed into other subroutines defined by the paper.
        Bottom: further decomposed into constant sized gates.
        Performs $\text{target} \rightarrow (\text{target} + d + \text{carry}_\text{in}) \bmod 2^n$.
    }
    \label{fig:add-3-clean}
\end{figure}
\begin{figure}
    \centering
    \begin{adjustwidth}{-2.5cm}{-2.5cm}
    \includegraphics[width=1.0\linewidth]{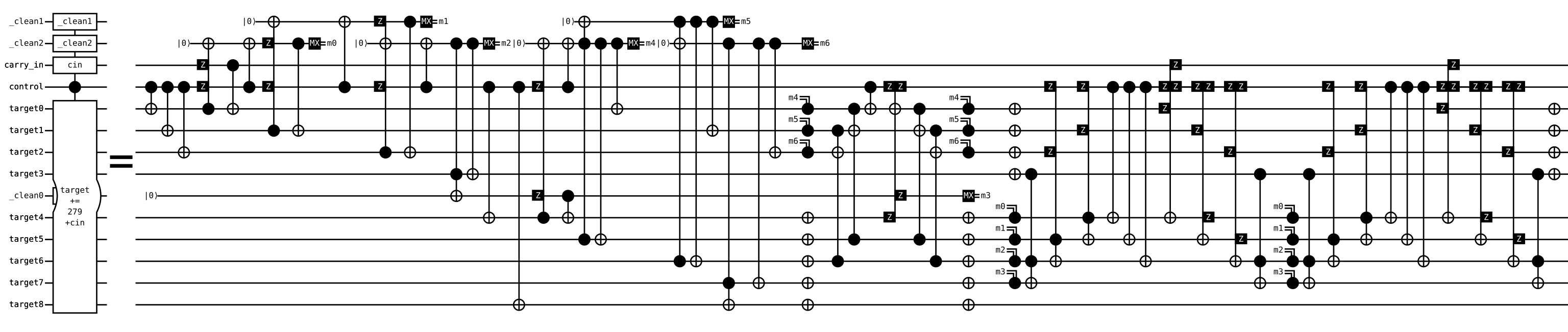}
    \end{adjustwidth}
    \caption{
    A controlled variant of \fig{add-3-clean}, specialized to the case $d=279$.
    Performs $\text{target} \rightarrow (\text{target} + \text{control} \cdot (279 + \text{carry}_\text{in})) \bmod 2^n$.
    279 in binary is 100010111, so this circuit substitutes $d_8, d_4, d_2, d_1, d_0 \rightarrow \text{control}$ and $d_7, d_6, d_5, d_3 \rightarrow 0$.
    }
    \label{fig:add-3-clean-controlled}
\end{figure}

\subsection{Borrowing Back and Forth}

In \cite{haner2016factoring}, a key idea was to split the addition into two halves.
I will similarly split the addition into two halves but, instead of recursively dividing each half into quarters and so forth, I intend to apply \fig{add-2-clean-n-dirty} to each half while borrowing the other as dirty qubits.
The steps are as follows:

\begin{enumerate}
\item
Perform a vented addition into the bottom half of the target register (the least significant half).
Crucially, this addition needs to compute the carry qubit needed for the top half addition.
Note that a carry-xor \emph{can't} be merged into the bottom half addition, because it would need to borrow the top half now but the top half needs to be unperturbed for the next step.
\item
Perform a vented addition into the top half.
This addition does include a merged carry-xor, borrowing the bottom half as the dirty target.
\item
Fix phases vented by the top half addition, by targeting the bottom half with a carry-xor conjugated by classically controlled Z gates.
The top half of the register is now finished.
\item
Vent the bottom-to-top carry qubit (produce a phasing task by measuring it in the X basis) now that it's no longer needed by the top half of the addition.
\item
Fix phases vented by the addition into the bottom half.
The top half is now available to be borrowed, and so can be targeted by two carry-xors interleaved with classically controlled Z gates.
This completes the entire addition.
\end{enumerate}

This construction uses 3 clean qubits (two for the vented additions and one to hold the bottom-to-top carry qubit) and $4n \pm O(1)$ Toffolis ($n \pm O(1)$ for the streaming additions and $3n \pm O(1)$ for carry-xors).
An example circuit is shown in \fig{add-3-clean}.

An interesting property of the adders presented in this paper is that they can all be controlled at no additional Toffoli cost.
To control an addition by an offset of $d$, replace every instance of $d_k$ where $d_k=0$ with $0$ and every instance of $d_k$ where $d_k=1$ with a usage of the control qubit.
Because $d_k$ is only ever used to invert controls and to control bit flips, this replacement introduces no new Toffoli gates.
An example is shown in \fig{add-3-clean-controlled}.
Another nice property all the shown circuits have is that the carry input qubit is only ever used as a control, meaning it can easily be replaced by a bit or removed entirely.

To generate other instances of the addition circuits shown in this paper, I refer readers to the Python code in the Zenodo upload~\cite{gidneyzenodo2025constadd}.

\section{Conclusion}
\label{sec:conclusion}

In this paper, I used ``venting'' (transforming unwanted qubits into unwanted phasing tasks by X basis measurement) to construct space-efficient quantum-classical adders.
I showed that a quantum-classical addition could be performed using $3n \pm O(1)$ Toffolis when assisted by $n-2$ dirty qubits and $2$ clean qubits, or with $4n \pm O(1)$ Toffolis when assisted by $3$ clean qubits.
I also noted that these constructions can be controlled at no additional Toffoli cost.

One downside of the constructions in this paper is their linear depth.
This is somewhat inherent to the goal of using constant workspace, since in practice applying operations in parallel requires workspace proportional to the parallelism to accommodate routing and magic production.
I leave finding low depth tradeoffs as future work for others to attempt.
It would be particularly interesting if venting could be used to finally create a true quantum carry-save adder; something able to accumulate offsets with $O(1)$ marginal depth while using $O(n)$ storage.
Existing attempts at carry-save adders aren't "true" because they have an $\Omega(\lg n)$ or $\Omega(\lg \lg 1/\epsilon)$ multiplicative overhead on the storage or the depth~\cite{gosset1998carrysave,draper2000qftadder,gidney2019approximate,kim2025carrysavetree,kim2025catalyticrotation}.

I'll close the paper by noting some personal significance.
My first research contribution was a linear-cost constant-workspace quantum incrementer~\cite{gidney2015largeincrement}.
I've wanted to know how to extend it to arbitrary offsets for \emph{10 years}.
In fact, based on private conversations, this problem was a bugbear for many researchers who work on quantum arithmetic circuits (because it appears in other tasks, like computing superposed multiplicative inverses).
In hindsight, the construction seems so simple.
I can't explain why it took us so long to find it.
I'm just glad we finally have it.

\section{Acknowledgments}

I thank Thiago Bergamaschi and Tanuj Khattar for providing comments which improved the paper.
I thank Hartmut Neven, and the entire Google Quantum AI team, for creating an environment where this work was possible.

\printbibliography

\end{document}